\documentclass{elsart}
\usepackage{graphicx}    
\usepackage{amssymb}

 \newcommand{\dtot}[2] { \frac{d {#1} } {d {#2}} }

 \newcommand{\text}[1] {\mbox{#1}}

 \newcommand{\eqref}[1] {(\ref{#1})}

\begin{document}

\begin{frontmatter}

\title{Dissipative perturbations for the $K(n,n)$ Rosenau-Hyman equation}

\author{Julio Garral\'on},
 \ead{jgarralon@uma.es}
\author{Francisco R. Villatoro\corauthref{COR}}
 \ead{villa@lcc.uma.es}

 \corauth[COR]{Corresponding author:
 Tel.: +34-951952388; fax: +34-951952542.}
\address{E.T.S. Ingenier{\'\i}a Industrial,
 Dept. Lenguajes y Ciencias de la Computaci\'on, \\
Universidad de M\'alaga, Escuela de Ingenier{\'\i}as, \\
C/ Doctor Ortiz Ramos, s/n, 29071 M\'alaga, Spain}



\begin{abstract}
Compactons are compactly supported solitary waves for nondissipative evolution equations with nonlinear dispersion. In applications, these model equations are accompanied by dissipative terms which can be treated as small perturbations. We apply the method of adiabatic perturbations to compactons governed by the $K(n,n)$ Rosenau-Hyman equation in the presence of dissipative terms preserving the ``mass" of the compactons. The evolution equations for both the velocity and the amplitude of the compactons are determined for some linear and nonlinear dissipative terms: second-, fourth-, and sixth-order in the former case, and second- and fourth-order in the latter one. The numerical validation of the method is presented for a fourth-order, linear, dissipative perturbation which corresponds to a singular perturbation term.
\end{abstract}


\end{frontmatter}

\section{Introduction}
\label{Intro}

Generalized Korteweg--de Vries equations with nonlinear dispersion can propagate compactly
supported solitary waves, referred to as compactons~\cite{RosenauHyman1993,CooperShepardEtAl1993,KhareCooper1993,Rosenau1994,Rosenau1996,OlverRosenau1996,Rosenau1997}. Numerical simulations show that an initial pulse wider than a compacton
decomposes into a set of compactons with a small amount of radiation; moreover, compactons
collide elastically suffering only a phase shift after the collision and generating a
small-amplitude, zero-mass, compact ripple~\cite{RosenauHyman1993,deFrutosSanzSerna1995,RusVillatoro2007,MihailaEtAl2010,CardenasEtAl2011}. First discovered in the (focusing) $K(n,n)$ Rosenau--Hyman equation for the modelling of pattern formation in liquid drops, compactons have several applications in physics and science~\cite{RosenauHyman1993},
such the pattern formation on liquid surfaces~\cite{Ludu1998}, nonlinear excitations in
Bose-Einstein condensates~\cite{Kovalev1998}, the lubrication approximation in thin films~\cite{Bertozzi1996}, or even
the pulse propagation in ventricle-aorta system~\cite{Kardashov2006}. The $K(n,n)$ equation is also the continuous limit of the discrete equations of a nonlinear lattice~\cite{RosenauHyman1993,Kivshar1993} and has been generalized to higher dimensions~\cite{Rosenau2007}. Let us also note that the $K(m,n)$ equation has also other kind of solutions, such as elliptic
compactons~\cite{RosenauHyman1993,Rosenau2000,Cooper2006}, loop solutions~\cite{ZhangLi2011},
dark solitons~\cite{TrikiWazwaz2009}, peakons~\cite{BinQing2010} and other solitary
waves~\cite{Biswas2008,EbadiBiswas2011}. Finally, let us remark that several generalizations of the $K(m,n)$ equation have also been considered in the literature, for example, the inclusion of time-dependent damping and dispersion~\cite{Biswas2010}, or the addition of fifth-order dispersion~\cite{CooperHymanKhare2001}.

In applications, one-dimensional nonlinear evolution equations with
solitary waves  are usually obtained as the leading order term of a
perturbative or asymptotic expansion of the solution of a more
complicated mathematical model~\cite{KevorkianCole,KivsharMalomed1989}. The perturbation is adiabatic when the ``mass" associated to the solitary wave is conserved~\cite{PertCero,LambBook,PertUno,PertDos}; in the context of solitons, such adiabatic perturbation methods are referred to as soliton perturbation theory~\cite{AntonovaBiswas2009,LailaBiswas2010,JohnsonBiswas2011}. Solitary waves are usually robust to adiabatic perturbations, even if they are singular, when their effect is weak; in such a case, the parameters of solitons and compactons, namely the velocity and the amplitude, slowly change in time.
Let us note that the $K(n,n)$ equation does not have a Lagrangian, hence the robustness of a compacton under perturbations cannot be studied by using the Lyapunov stability method developed by Dey and Khare~\cite{DeyKhare1998} for the Cooper--Shepard--Sodano equation~\cite{CooperShepardEtAl1993,KhareCooper1993}.

Adiabatic perturbations methods have been applied to the compactons of the $K(2,2)$ equation with second- and fourth-order linear dissipative perturbations in Ref.~\cite{Pikovsky2006} and numerical simulations show their good accuracy in determining the evolution of the velocity and amplitude of the perturbed compactons~\cite{RusVillatoro2009b}. The damping of the amplitude of the compacton is accompanied by the generation of trailing tails resulting from the conservation of its ``mass"~\cite{RusVillatoro2009a,AbassyEtAl2009,RusVillatoro2010}. This paper extends the adiabatic perturbation theory to compactons of the $K(n,n)$ equation under weak dissipative perturbations. Next section recalls the main properties of the perturbed $K(n,n)$ equation. Section~\ref{APTC} applies the adiabatic perturbation method to five dissipative perturbation terms. The validity of the method for singular perturbations is illustrated by using numerical solutions in Section~\ref{Tails:validation}. Finally, the last section is devoted to the main conclusions.

\section{The perturbed $K(n,n)$ equation}\label{PKnn}

Let us consider the $K(m,n)$ equation given by
\begin{equation}
 \label{Knm}
  {u}_{t} + (u^m)_{x} + (u^n)_{xxx} = 0,
\end{equation}
where $u(x,t)$ is the wave amplitude, $x$ is the spatial coordinate, $t$ is time, and the subindex indicate differentiation. This equation has at least two invariants given by
\begin{equation}
 \label{Knm:M:P}
 M\equiv I_1 = \int u\,dx, \qquad \text{and}\qquad P\equiv I_2 = \int u^{n+1}\,dx,
\end{equation}
referred to as ``mass" and ``momentum" in analogy with the $K(m,1)$ equation which has another invariant given by
\[
 E\equiv I_3= \int \left( \frac{(u_x)^2}{2}-\frac{u^{m+1}}{m+1}\right)\,dx,
\]
referred to as ``energy"~\cite{DrazinBook}. Some authors refer to $P$ as energy in the case of $n=m$, but in our opinion momentum is more appropriate.

The perturbed $K(n,n)$ equation is given by
\begin{equation}
 \label{Kpp:pert}
  {u}_{t} + (u^n)_{x} + (u^n)_{xxx} = \varepsilon\,\mathcal{P}(u),
\end{equation}
where $|\varepsilon|\ll 1$ is a small parameter and $\mathcal{P}(u)$
is a function of $u$ and its spatial and temporal derivatives. Under weak perturbations, the robustness of compactons results in the preservation of their shape only with a slow time evolution of their parameters; in such a case, by introducing a slow time $\tau=\varepsilon\,t$, the solution of Eq.~\eqref{Kpp:pert} can be written as
\begin{equation}
 \label{compacton:plus:tail}
 u(x,t,\tau) = u_c(x,t,\tau) + u_T(x,t,\tau),
\end{equation}
where $u_c$ corresponds to a compacton whose parameters evolve slowly in time and $u_T$ is the non-compacton part of the solution, usually a trailing tail. The evolution of the parameters of the compacton $u_c(x,t,\tau)$ of Eq.~\eqref{Kpp:pert} corresponds to the ansatz given
by
\begin{equation}
 \label{compactonsol2}
 u_c(x,t,\tau) =
  \left\{\frac{2\,n\,c(\tau)}{(n+1)}\,
  \cos^{2} \left(
        \frac{n-1}{2\,n}\,\xi
       \right)\right\}^{{1}/{(n-1)}},
\end{equation}
for $|\xi|\le {n\,\pi}/({n-1})$, and $u_c(x,t,\tau) = 0$ otherwise, where  $\xi\equiv\xi(x,t,\tau)=x-c(\tau)\,t$; hence the compacton's amplitude is $\{{2\,n\,c(\tau)}/{(n+1)}\}^{{1}/{(n-1)}}$, and its velocity is $c(\tau)$. Let us remark that the solution~\eqref{compactonsol2} of Eq.~\eqref{Kpp:pert} with $\varepsilon=0$ was first obtained for $n=2$ and~$3$ in Ref.~\cite{RosenauHyman1993}, and generalized in Ref.~\cite{KhareCooper1993} for $n\in (1,3)$; the last condition is required for Eq.~\eqref{compactonsol2} to be a solution in the classical sense. Let us also note that the derivation of both invariants in Eq.~\eqref{Knm:M:P} is independent of the number of continuous derivatives at both edges of the compacton~\eqref{compactonsol2} for $n\in(1,3)$.

For perturbations which do not preserve the mass of the compacton, the evolution in the slow time of the mass can be used to determine the evolution of $c(\tau)$. The integration in space of Eq.~\eqref{Kpp:pert} yields
\begin{equation}
 \label{firstInvPert}
  \frac{d}{d\tau} \int_{-\infty}^{\infty} {u}\,dx
  = \varepsilon\,\int_{-\infty}^{\infty} \mathcal{P}(u)\, dx.
\end{equation}
Taking $\varepsilon\,\mathcal{P}(u) = -\varepsilon_0\,u$,
with $|\varepsilon_0|\ll 1$, and introducing
$u_c$ given by Eq.~\eqref{compactonsol2} into Eq.~\eqref{firstInvPert} yields
\[
 \dtot{c(\tau)}{\tau} =  -(n-1)\,c(\tau),
\]
whose solution is
\begin{equation}
 c(\tau) = c(0)\,e^{-\varepsilon_0\,(n-1)\,\tau},
 \label{sol:ctau:alpha0}
\end{equation}
indicating that both the velocity and the amplitude of the compacton decrease slowly as it evolves in time for $n>1$.

\section{Adiabatic perturbation theory for compactons}\label{APTC}

The adiabatic perturbation theory is a technique for the analysis of solitary wave solutions of nonlinear evolution equations under dissipative perturbations preserving the mass of the compacton (let's notice that in such a case the right-hand side of Eq.~\eqref{firstInvPert} is null). This technique determines the slow time evolution of the parameters of the compacton by means of using the slow temporal derivative of the invariant of Eq.~\eqref{Kpp:pert} with
$\varepsilon=0$. Multiplying Eq.~\eqref{Kpp:pert} by $u^n$ and
integrating in space results in
\begin{equation}
  \frac{d}{d\tau} \int_{-\infty}^{\infty} \frac{u^{n+1}}{n+1}\,dx
  = \int_{-\infty}^{\infty} u^n\,\mathcal{P}(u)\, dx,
 \label{secondInvPert}
\end{equation}
where the substitution of $u$ by $u_c$ in Eq.~\eqref{compactonsol2}
gives an ordinary differential equation for the velocity $c(\tau)$
as function of the slow time $\tau$. In such a case, the integral of the
left-hand side of Eq.~\eqref{secondInvPert} yields
\begin{eqnarray}
 &&\int_{-\infty}^{\infty} \frac{(u_c)^{n+1}}{n+1}\,dx
  = \frac{\sqrt{\pi }}{(n+1)}\,\left(\frac{2\,n}{n+1}\right)^{\frac{2\,n}{n-1}}\,
    \frac{\Gamma \left(\frac{3\,n+1}{2\,(n-1)}\right)}{ \Gamma
    \left(\frac{n+1}{n-1}\right)}\,
    c(\tau)^{\frac{n+1}{n-1}}, \nonumber
\end{eqnarray}
and its right-hand side results in a similar expression depending of the perturbation $\mathcal{P}(u)$ considered.

Next subsections illustrate this  method by using several dissipative perturbations, both linear and nonlinear. Other perturbations not considered in this paper may be dealt with
similarly.

\subsection{Perturbation with second-order derivatives}
\label{Pert2ndOrder}

Let us study the effect of a linear perturbation given by
\begin{equation}
 \varepsilon\,\mathcal{P}(u)=
   \alpha_0\,u_{xx}
  -\alpha_1\,u_{xt}
  +\alpha_2\,u_{tt}
 ,
   \label{2nd:pert}
\end{equation}
where $0<\alpha_i\ll 1$. By means of straightforward integration
using standard mathematical software, Eq.~\eqref{secondInvPert}
gives
\begin{eqnarray}
 &&
 c'(\tau) =
   - \frac{2\,(n-3)}{n\,(n+3)}
   \frac{\Gamma\left(\frac{3-n}{n-1}\right)}
        {\Gamma\left(\frac{3-n}{n-1}+2\right)}
   \,
    \left(
   \alpha_0\,c(\tau)
   +\alpha_1\,c(\tau)^2
   +\alpha_2\,c(\tau)^3
   \right)
   .
\end{eqnarray}
By using the properties of the Gamma function, for $1<n<3$, this
equation can be simplified to yield
\begin{equation}
 c'(\tau) =
   - \frac{(n-1)^2}{n\,(n+3)}
   \,\left(
   \alpha_0\,c(\tau)
   +\alpha_1\,c(\tau)^2
   +\alpha_2\,c(\tau)^3
   \right)
   .
   \label{ode:2nd:pert}
\end{equation}
The analytical solution of this equation can be obtained in implicit
form, for $\alpha_2=0$ also in explicit form, showing that the
perturbation~\eqref{2nd:pert} is dissipative for $\alpha_i>0$.

\subsection{Perturbation with fourth-order derivatives}
\label{Pert4thOrder}

Let us take a fourth-order linear perturbation given by
\begin{equation}
 \varepsilon\,\mathcal{P}(u)=
   -\beta_0\,u_{xxxx}
   +\beta_1\,u_{xxxt}
   -\beta_2\,u_{xxtt}
   +\beta_3\,u_{xttt}
   -\beta_4\,u_{tttt}
 ,
   \label{4th:pert}
\end{equation}
where $0<\beta_i\ll 1$. After a long integration process of
Eq.~\eqref{secondInvPert} gives
\begin{eqnarray*}
 &&
 c'(\tau) =
   - \frac{(n-1)^3\,((n-3)\,n-1)}{(n-5)\,n^3\,(n+3)}
   \,
   \left(
   \beta_0\,c(\tau)
   +\beta_1\,c(\tau)^2
   +\beta_2\,c(\tau)^3
   \right.
 \\&& \phantom{c'(\tau) = - \frac{(n-1)^3\,((n-3)\,n-1)}{(n-5)\,n^3\,(n+3)} \qquad }
   \left. \phantom{1}
   +\beta_3\,c(\tau)^4
   +\beta_4\,c(\tau)^5
   \right),
\end{eqnarray*}
for $1<n<3$. The analytical solution of this equation can be
obtained in implicit form and their numerical evaluation shows that
the perturbation~\eqref{4th:pert} is dissipative for $\beta_i>0$.

\subsection{Perturbation with sixth-order derivatives}
\label{Pert6thOrder}

The last linear perturbation to be considered in this paper is
\begin{eqnarray*}
 &&
 \varepsilon\,\mathcal{P}(u)=
   \gamma_0\,u_{xxxxxx}
  -\gamma_1\,u_{xxxxxt}
  +\gamma_2\,u_{xxxxtt}
  -\gamma_3\,u_{xxxttt}
 \\ &&
 \phantom{\varepsilon\,\mathcal{P}(u)=1}
  +\gamma_4\,u_{xxtttt}
  -\gamma_5\,u_{xttttt}
  +\gamma_6\,u_{tttttt}
 ,
\end{eqnarray*}
where $0<\gamma_i\ll 1$. For this perturbation,
Eq.~\eqref{secondInvPert} yields
\begin{eqnarray*}
 &&
 c'(\tau) =
   -
   \frac{(n-1)^4\,\left(3\,n^4-13\,n^3+7\,n^2+13\,n+5\right)}
        {(n-5)\,n^5\,(n+3)\,(3\,n-7)}\left(
    \gamma_0\,\,c(\tau)+ \gamma_1\,\,c(\tau)^2
   \right.
  \,
 \\ &&
 \phantom{c'(\tau)= \qquad}
   \left. \phantom{1}
  + \gamma_2\,\,c(\tau)^3
  + \gamma_3\,\,c(\tau)^4
  + \gamma_4\,\,c(\tau)^5
  + \gamma_5\,\,c(\tau)^6
  + \gamma_6\,\,c(\tau)^7
  \right),
\end{eqnarray*}
for $1<n< 7/3$; note that the integration of
Eq.~\eqref{secondInvPert} for $n\ge 7/3$ does not converge to a
finite value. The analytical solution of this equation can be
obtained in implicit form and their numerical evaluation shows that
the perturbation~\eqref{4th:pert} is dissipative for $\gamma_i>0$.

\subsection{Nonlinear perturbation with second-order derivatives}
\label{PertNL2ndOrder}

Dissipative perturbations can also be nonlinear, for example, the
second-order nonlinear perturbation given by
\begin{equation}
 \varepsilon\,\mathcal{P}(u)=
  \delta_1\,(n-1)\,n\,(u_x)^2\,u^{n-2}
 +\delta_2\,n\,u_{xx}\,u^{n-1}
 ,
   \label{nl:2nd:pert}
\end{equation}
where $0<\delta_i\ll 1$. Note that
$\varepsilon\,\mathcal{P}(u)=\delta\,(u^n)_{xx}$ for $\delta_1=\delta_2=\delta$.
By means of straightforward integration, in this case, Eq.~\eqref{secondInvPert}
gives
\begin{equation}
 c'(\tau) =
  (\delta_1\,(n-1)+\delta_2\,(1-2\,n))
  \,\frac{(n-1)^2}{2\,n\,(n+1)}
  \,c(\tau)^2
   .
   \label{nl:2nd:pert:ode}
\end{equation}
This equation reduces to
\begin{equation}
 c'(\tau) =
  -\delta\,\frac{(n-1)^2}{2\,(n+1)}
  \,c(\tau)^2
   ,
   \label{nl:2nd:pert:ode:all}
\end{equation}
for $\delta_1=\delta_2=\delta$.

The analytical solution of Eq.~\eqref{nl:2nd:pert:ode:all} shows
that the perturbation~\eqref{nl:2nd:pert} is dissipative for
$\delta>0$, but that of Eq.~\eqref{nl:2nd:pert:ode} shows that a
dissipative perturbation requires, for $\delta_2=0$, that
$\delta_1<0$ for $1<n<2$, or $\delta_1>0$ for $2<n<3$, and, for
$\delta_1=0$, that $\delta_2>0$ for $1<n<3/2$ or $2<n<3$, or
$\delta_2<0$ for $3/2<n<2$.

\subsection{Nonlinear perturbation with fourth-order derivatives}
\label{PertNL4thOrder}

Our final example of a nonlinear dissipative perturbation is
\begin{eqnarray*}
 &&
 \varepsilon\,\mathcal{P}(u)=
  \eta_1\,(u_x)^4\,u^{n-4}
 +\eta_2\,(u_x)^2\,u_{xx}\,u^{n-3}
 +\eta_3\,(u_{xx})^2\,u^{n-2}
 \\ && \phantom{ \varepsilon\,\mathcal{P}(u)= \qquad 1}
 +\eta_4\,u_x\,u_{xxx}\,u^{n-2}
 +\eta_5\,u_{xxxx}\,u^{n-1}
 ,
   \label{nl:4th:pert}
\end{eqnarray*}
where $0<\eta_i\ll 1$. After a long integration,
Eq.~\eqref{secondInvPert} yields
\begin{eqnarray}
 &&
 c'(\tau) =
  \frac{(n-1)^3\,c(\tau)^2}{2\,n^4\,(n+1)\,(n+3)}
  \,
  \left(
    3\,\eta_1-\eta_2\,(2\,n-3)+\eta_3\left(2\,n^2-2\,n+3\right)
  \right. \nonumber \\&& \phantom{c'(\tau) = 1}
 \nonumber \\ &&
 \phantom{c'(\tau)=\quad 1}
  \left.
    +\eta_4\,\left(2\,n^2-8\,n+3\right)
    -\eta_5\,(2\,n-1)\,\left(2\,n^2-8 \,n+3\right)
  \right)
   .
   \label{nl:4th:pert:ode}
\end{eqnarray}
The conditions on $\eta_i$ such that the solution of this equation
is dissipative must be discussed in a term-by-term basis with a
$\eta_i\ne 0$ and the other ones $\eta_j=0$, for $j\ne i$. The first
term is dissipative for $\eta_1<0$; the second one, for $\eta_2<0$
with $1<n<3/2$, and for $\eta_2>0$ with $3/2<n<3$; the third one,
for $\eta_3<0$; the fourth one, for $\eta_4>0$; and the fifth one,
for $\eta_5<0$. Further possibilities require a straightforward but cumbersome combination of these facts.

Note that for $\varepsilon\,\mathcal{P}(u)=\eta\,(u^n)_{xxxx}$, i.e. for
$\eta_1=\eta\,(n-3)\,(n-2)\,(n-1)\,n$,
$\eta_2=6\,\eta\,(n-2)\,(n-1)\,n$,  $\eta_3=3\,\eta\,(n-1)\,n$,
$\eta_4=4\,\eta\,(n-1)\,n$, and $\eta_5=\eta\,n$,
Eq.~\eqref{nl:4th:pert:ode} reduces to
\begin{equation}
 c'(\tau) =
  \eta\,\frac{(n-1)^3\,(2+n)\,c(\tau)^2}{2\,n\,(n+1)\,(n+3)},
   \label{nl:4th:pert:ode:all}
\end{equation}
whose analytical solution it dissipative for $\eta<0$.

\section{Validation for singular perturbations}
\label{Tails:validation}

The main criticism on the validity of the adiabatic perturbation method presented in the last section is the question of its applicability to singular perturbations, i.e., when $\varepsilon\,\mathcal{P}(u)$ has more than three spatial derivatives. Such terms may cause a fundamental change at the edge of the compacton and may invalidate the whole structure. Let us check by numerical simulation that, for small enough $\varepsilon$, the method works properly.

The numerical solution of the perturbed $K(n,n)$ equation with a fourth-order linear, dissipative term given by
\begin{equation}
 \label{Kpp:pert:lin:cuarto}
  {u}_{t} -c_0\,u_x + (u^n)_{x} + (u^n)_{xxx} + \beta_0\,u_{xxxx} = 0,
\end{equation}
can be obtained by the Petrov-Galerkin method presented in Refs.~\cite{deFrutosSanzSerna1995,RusVillatoro2009b,MihailaEtAl2010}, using the implicit midpoint rule for the integration in time~\cite{RusVillatoro2010}; a detailed presentation is omitted here for the sake of brevity.

The perturbed compacton~\eqref{compacton:plus:tail} develops a trailing tail under the dissipative term, but evolves preserving the mass of the solution, cf. Eq.~\eqref{Knm:M:P}, hence
\begin{equation}
 \label{compacton:plus:tail:mass}
 M = \int u(x,t,\tau)\,dx = \int u_c(x,t,\tau)\,dx + \int u_T(x,t,\tau)\,dx.
\end{equation}
From this expression the area under the tail can be easily calculated yielding
\begin{eqnarray}
 &&
 A_T(\tau) = \int u_T(x,t,\tau)\,dx = M - \int u_c(x,t,\tau)\,dx
 \nonumber \\ && \phantom{A_T(t) }
        = n\,\sqrt{\pi}\,\left(\frac{2^n\,n}{n+1}\right)^{1/(n-1)}\,\frac{\Gamma(1/2+1/(n-1))}{\Gamma(1/(n-1))}\,
 \nonumber \\ && \phantom{A_T(t) = 1 }
          \left( c(0)^{1/(n-1)} - c(\tau)^{1/(n-1)}
          \right).
 \label{compacton:plus:tail:area}
\end{eqnarray}

\begin{figure}
\centering
{\includegraphics[width=7cm]{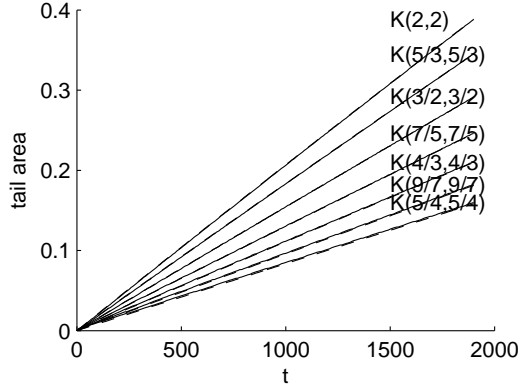} }
\caption{Evolution in time of the area under the tail of a compacton calculated by a quadrature rule applied to the numerical solution (solid line) and with an adiabatic approximation (dashed one) for the $K(n,n)$ equation with $n = 2$, 5/3, 3/2, 7/5, 4/3, 9/7, and $5/4$. The compacton has a velocity $c=1$, and the numerical method uses $c_0=1$, $\beta_0=0.001$, $\Delta x = 0.2$, and $\Delta t = 0.1$.}
\label{fig:colas:quad}
\end{figure}

Figure~\ref{fig:colas:quad} shows the evolution in time of the area under the tail of a compacton calculated by a quadrature rule applied to the numerical solution of Eq.~\eqref{Kpp:pert:lin:cuarto} (solid line) and by using Eq.~\eqref{compacton:plus:tail:area} (dashed one) for the $K(n,n)$ equation with $n = 2$, 5/3, 3/2, 7/5, 4/3, 9/7, and $5/4$. The plot shows the good accuracy of the adiabatic perturbation method even for a singular perturbation.

Equation~\eqref{compacton:plus:tail:area} can be used to estimate the shape of the trailing tail, except at its front. The left edge of the perturbed compacton is located at position $X(\tau)$ given by
\[
 X(\tau) = X(0)+\int_0^\tau c(z)\,dz,
\]
where $X(0)$ is the location of the  left edge of the compacton at $t=0$. Hence, the tail area~\eqref{compacton:plus:tail:area} can be calculated as
\[
 A_T(\tau) = \int_{X(0)}^{X(\tau)}  u_T(x,t,\tau)\,dx.
\]
The shape of the tail can be obtained by using Leibniz's rule for differentiation under the integral sign, yielding
\begin{equation}
 \label{compacton:tail:estimation}
  u_T(X(\tau),t,\tau) = \frac{1}{c(\tau)}\,\dtot{ A_T(\tau) }{\tau}.
\end{equation}

\begin{figure}
\centering
{\includegraphics[width=14cm]{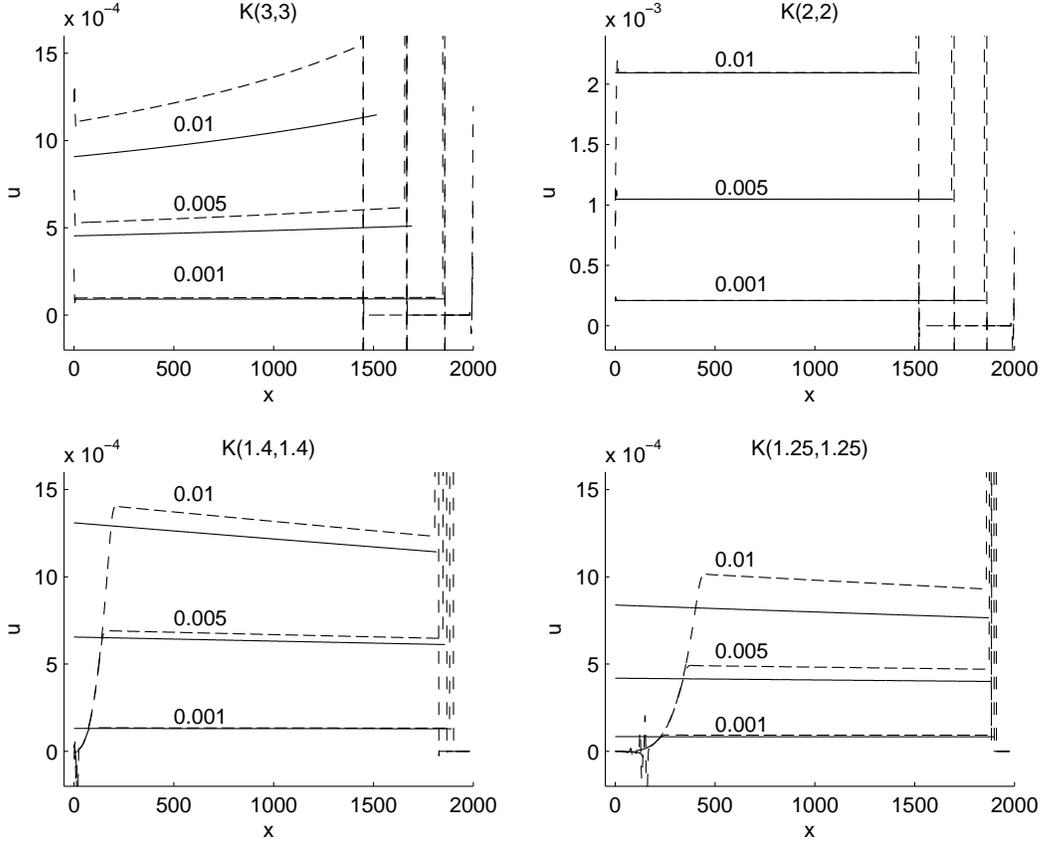} }
\caption{The shape of the tails of a compacton obtained by means of the adiabatic perturbation method (solid line) and calculated by using a numerical method (dashed one) for the $K(n,n)$ equation with fourth-order linear dissipations $\beta_0=0.01$, 0.005, and 0.001, for $n = 3$ (left top plot), 2 (right top one), 7/5 (left bottom one), and 5/4 (right bottom one). In all the plots the compacton has a velocity $c=1$, $t=1900$, and the numerical method uses $c_0=1$, $\Delta x = 0.2$, and $\Delta t = 0.1$.}
\label{fig:areas:quad}
\end{figure}

Figure~\ref{fig:areas:quad} shows the shape of the tails of a perturbed compacton obtained by means of Eq.~\eqref{compacton:tail:estimation} (solid line) and numerically calculated by using Matlabs's ODE suite (dashed one) for Eq.~\eqref{Kpp:pert:lin:cuarto} with $n = 3$ (left top plot), 2 (right top one), 7/5 (left bottom one), and 5/4 (right bottom one); each plot presents three curves for $\beta_0=0.01$, 0.005, and 0.001. The plots zoom into the tail because of its small amplitude compared to that of the compacton. For the $K(2,2)$ (right top plot) the agreement between the adiabatic perturbation method and the numerical result is extremely good due to the small width of the front of the trailing tail. As the value of $n<2$ approaches unity from above, the width of the front of the tail increases, resulting in a loss of the accuracy of Eq.~\eqref{compacton:tail:estimation} as shown in the bottom plots in Fig.~\ref{fig:areas:quad}; even in such a case, for small enough values of the parameter $\beta_0$, the agreement between perturbation and numerical results is good. For the limiting case $n=3$ (left top plot in Fig.~\ref{fig:areas:quad}), the accuracy of the adiabatic perturbation method developed in this paper is worst except for small values of $\beta_0$.

\section{Conclusions}
\label{Conclusions}

The adiabatic perturbation method has been applied to the $K(n,n)$ Rosenau-Hyman equation with both linear and nonlinear dissipation terms. The slow time dynamics of the compacton velocity (which uniquely determines that of its amplitude) has been determined for five different perturbations. The analytical results have been validated by means of numerical methods for a singular perturbation, showing the good accuracy of the adiabatic perturbation method, even for the estimation of the shape of the trailing tails of the perturbed compactons.

The adiabatic perturbation method can be applied to other nonlinear evolution equations under dissipative perturbations having compacton solutions, such the $K^*(n,n)$ Cooper-Shepard-Sodano equation, and generalized versions of the Boussinesq, regularized long-wave, Benjamin-Bona-Mahony, and Camassa-Holm equations, to mention only a few~\cite{RusVillatoro2009a}.

\section*{Acknowledgments}

The research reported here was partially supported by Projects
MTM2010--19969 (F.R.V.) and TIN2008--05941 (J.G.) from the
Ministerio de Ciencia e Innovaci\'on of Spain, and Project TIC-6083
(J.G.) from the Junta de Andaluc\'{\i}a.

\newcommand{\paperref}[8]{{#1}, {#2}, {#8} {#4} ({#7}) {#6}.}

\newcommand{\bookref}[5]{{#1}, {#2}, {#3}, {#4}, {#5}}

\end{document}